\documentclass[universe,article,accept,moreauthors,pdftex]{Definitions/mdpi}

\usepackage{amsmath,amssymb}
\firstpage{1} 
\pubvolume{1}
\issuenum{1}
\articlenumber{0}
\pubyear{2021}
\copyrightyear{2021}
\externaleditor{Academic Editor: Sergio Cristallo, Paolo Ventura}
\datereceived{} 
\dateaccepted{} 
\datepublished{} 
\hreflink{https://doi.org/} 

\Title{Dust Production around Carbon-Rich Stars: The Role \mbox{of Metallicity}}

\TitleCitation{Dust Production around Carbon-Rich Stars: The Role of Metallicity}

\Author{Ambra Nanni 
 $^{1,}$*\orcidA, Sergio Cristallo $^{2,3}$\orcidB, Jacco Th. van Loon $^{4}$\orcidC ~and Martin A. T. Groenewegen $^{5}$\orcidD}

\AuthorNames{Ambra Nanni, Sergio Cristallo, Jacco Th. van Loon and Martin A. T. Groenewegen}

\AuthorCitation{Nanni, A.; Cristallo, S.; van Loon, J.T.; Groenewegen, M.A.T.}

\address{%
$^{1}$ \quad National Centre for Nuclear Research, ul. Pasteura 7, 02-093 Warszawa, Poland\\
$^{2}$ \quad INAF, Oservatorio Astronomico d'Abruzzo, Via Mentore Maggini snc, 64100 Teramo, Italy; sergio.cristallo@inaf.it 
\\ 
$^{3}$ \quad INFN, Sezione di Perugia, Via A. Pascoli snc, 06123 Perugia, Italy \\
$^{4}$ \quad Lennard-Jones Laboratories, Keele University, Keele 
 ST5 5BG, U.K.; j.t.van.loon@keele.ac.uk\\
$^{5}$ \quad Koninklijke Sterrenwacht van Belgi\"e, Ringlaan 3, B-1180 Brussel, Belgium; martin.groenewegen@oma.be}

\corres{Correspondence: ambra.nanni@ncbj.gov.pl}

\abstract{Background: Most of the stars in the Universe will end their evolution by losing their envelope during the thermally pulsing asymptotic giant branch (TP-AGB) phase, enriching the interstellar medium of galaxies with heavy elements, partially condensed into dust grains formed in their extended circumstellar envelopes. Among these stars, carbon-rich TP-AGB stars (C-stars) are particularly relevant for the chemical enrichment of galaxies. We here investigated the role of the metallicity in the dust formation process from a theoretical viewpoint.
Methods: We coupled an up-to-date description of dust growth and dust-driven wind, which included the time-averaged effect of shocks, with FRUITY stellar evolutionary tracks. We compared our predictions with observations of C-stars in our Galaxy, in the Magellanic Clouds (LMC and SMC) and in the Galactic Halo, characterised by metallicity between solar and 1/10 of solar.
Results: Our models explained the variation of the gas and dust content around C-stars derived from the IRS Spitzer spectra. The wind speed of the C-stars at varying metallicity was well reproduced by our description. We predicted the wind speed at metallicity down to 1/10 of solar in a wide range of mass-loss rates.}


\begin{document}
\section{Introduction}
The thermally pulsing asymptotic giant branch (TP-AGB) phase represents the final stage in the evolution of low- and intermediate-mass stars (M$<(8-11)$~M$_\odot$, depending on the metallicity). During~the AGB phase, stars lose mass at high rates, up~to a few 10$^{-5}$~M$_\odot$~y$^{-1}$, enriching the interstellar medium (ISM) with elements freshly synthesised in their interior. 
While the role of such stars as dust producers remains controversial in high-redshift galaxies \citep{Valiante09, Dwek11, Michalowski15, Lesniewska19, Dellagli19}, they are important dust producers in the local Universe \citep{Gehrz89}.

The mass-loss mechanism of TP-AGB stars is initiated by pulsations, which induce shock waves and allow the gas to move outwards the stellar surface where the temperature is low enough to allow dust condensation. If~the momentum transferred by stellar photons incident on dust grains overcomes the gravity of the stars, the~star loses mass by means of a dust-driven wind \citep{Hoefner18}. In~addition to mass loss, the~TP-AGB phase is characterised by other complex physical processes, whose interplay determines the composition of dust and the structure of the dense and extended circumstellar envelope (CSE) around them. One of the most important is the third dredge-up (TDU), through which some of the carbon processed in the internal layers is brought up to the stellar surface. Such a mechanism follows an expansion of the stellar upper layers due to the occurrence of thermonuclear runaways triggered by the sudden energy release from 3$\alpha$ processes (the so-called thermal pulses (TPs)). 
The efficiency of TDU changes as a function of the stellar mass and metallicity and is related to the capability of the convective envelope to penetrate inside the star and bring up carbon-rich material to the surface (see, e.g.,~\citep{cris11,stra06}). The TDUs occurring after TPs dramatically modify the chemical and physical properties of the stellar atmosphere and CSEs, eventually leading to a change in their spectral type. As a matter of fact, stars are classified as O-rich if the number of oxygen atoms exceeds the number of carbon atoms (i.e., \mbox{C/O $<$ 1}) or C-rich if C/O $>$ 1. 
For stellar masses $\gtrsim 4$~M$_\odot$, the CNO cycle may be activated at the base of the convective envelope. Such a process is called ``hot bottom burning'' (HBB) through which carbon nuclei are converted into nitrogen ones \citep{ventura11, kakka}. The~effect of the HBB is therefore to decrease the number of carbon nuclei favouring the formation of oxygen-rich TP-AGB~stars.

Dust grains around O-rich TP-AGB stars are mainly composed of silicates and other oxides (e.g., alumina oxide, Al$_2$O$_3$), while amorphous carbon (amC), silicon carbide (SiC) and, possibly, magnesium sulphide (MgS) form around C-rich stars. Iron dust may also be condensed, but~its presence is not straightforwardly identifiable because such a species does not produce specific features \citep{Marini19a}.
The production of dust in O-rich TP-AGB stars and of SiC and MgS around C-stars is proportional to the metallicity. 
On the contrary, the~production of amC depends on the efficiency of the TDU mechanism. As a consequence, low-metallicity C-stars may be relevant dust factories in the early \mbox{Universe \citep{nanni13, Nanni14}}, since the production of amC dust is not explicitly linked to their initial metallicity. In the Solar System, silicon carbide (SiC) is found in meteoric presolar grains and provides important constraints to the formation of the Sun, which is believed to have originated from the ashes of dying C-stars with initial mass M$_{\rm ini}\approx$ 2~M$_\odot$ \citep{cris20}. SiC presolar grains are rarer than clusters of metal carbides (TiC and Zr-Mo carbide) \citep{Bernatowicz96}, which might have acted as seed nuclei for the formation of amC dust. An~alternative (or complementary) path for the formation of amC grains involves the growth of aromatic rings and the formation of polycyclic aromatic hydrocarbons (PAHs) \citep{che92, Cherchneff00}. The~accretion of amC grains can then further proceed through the addition of carbon atoms via surface reactions with acetylene molecules in the gas~phase.

The production of amC dust is responsible for the outflow acceleration and mass loss in C-stars, which has been extensively investigated in recent years from the observational and theoretical point of view (e.g., \citep{Mattsson10, Eriksson14, McDonal18, Bladh19, Sandin20}). The~wind speed measured around C-stars can provide an important indication in relation to a possible metallicity dependence of the dust condensation process, the outflow velocity being linked to the amount of dust produced, as well as to the stellar parameters. 
To date, in~contrast to the ample measurements that have been made in the Galaxy \citep{Loup97, Schoier13, Ramstedt14, Danilovich15}, such information has only been obtained for a handful of C-stars of low metallicity---three in the Halo and the Sgr dSph \citep{Groenewegen97, Lagadec10}, three in the thick disc \citep{Lagadec10} and~four in the LMC \citep{Groenewegen_etal16}. While they appear to confirm lower expansion velocities at lower metallicity, the~observations are yet too sparse and biased to provide a comprehensive picture.
From a theoretical viewpoint, hydrodynamical calculations have highlighted that similar mass-loss rates and expansion velocities are obtained at varying metallicity as long as the other stellar parameters are comparable. Indeed, amC dust, which is metallicity independent, is the driver of the wind (e.g., \citep{Bladh19, Mattsson08}).

Direct evidence of dust production around C-stars has been probed down to 1/50 of solar in local dwarf galaxies, a~metallicity that can be representative of high-redshift \mbox{galaxies \citep{Boyer15, Boyer17, Jones18, Goldman19}.}
In the Magellanic Clouds (MCs), the~population of evolved stars has been probed by different observing programs that have allowed a photometric coverage of evolved stars from the ultraviolet to midinfrared (e.g., \citep{Boyer11, Riebel12}), as~well as for the measurements of spectra in the optical and midinfrared for a sample of evolved stars \mbox{(e.g., \citep{Boyer15a, Ruffle15, Jones17}).} 

The increasing quality and spectral coverage of observations has allowed different authors to constrain the gas, as well as the dust production and composition around TP-AGB stars by employing different theoretical and interpretative tools, especially in our Galaxy and in the close-by MCs (e.g., \citep{vanLoon99, vanLoon05, Blommaert06, Lebzelter06, Groenewegen07, vanLoon06, Matsuura06, vanLoon08, Groenewegen09, Riebel12, Gullieuszik12, Sloan16, Rau17, Goldman17, Goldman18, Nanni18, Nanni19b, Brunner19, Marini19a, Marini19b, Marini20a, Marini20b, Groenewegen20, Ramstedt20}). 
Different studies have evaluated the dust production rate for the entire population of evolved stars in the MCs, showing that the population of C-stars provides the most relevant contribution to the present-day dust enrichment of these galaxies \citep{Matsuura09, Boyer12, Riebel12, Matsuura13, Dellagli15, Srinivasan16}, contrary to what has been found at solar and supersolar metallicity (e.g., \citep{Javadi13}).
The analysis of nearinfrared colours first provided important clues to understand the dependence of dust condensation on the metallicity \citep{vanloon00}.
More recently, studies of the midinfrared Spitzer IRS spectra of C-stars in our Galaxy and in the MCs have allowed for a more complete characterisation of the gas and dust content, which has been investigated in a series of papers \citep{Matsuura06, Buchanan06, Sloan06b, Zijlstra06, Lagadec07, Leisenring08, vanLoon08, Kemper10, Sloan16}.
Specifically, by~analysing the midinfrared spectral features, it has been possible to constrain the variation of the acetylene (C$_2$H$_2$) content in the gas phase, a~molecule that constitutes the building blocks of amC dust grains, of~SiC and MgS production as a function of the mass-loss rate and metallicity. Despite the availability of observational data, the~process of dust formation in the CSEs of C-stars remains controversial and difficult to assess given the interplay among stellar evolution, dust formation and mass loss, as~well as our limited knowledge on the details of the microphysics of dust~grains.

In this paper, we studied the role of metallicity (and metallicity-related stellar properties) in the dust condensation process, and we derived some theoretical quantities that can be constrained by available and future observations, such as the C$_2$H$_2$ content, the~SiC fraction and the expansion velocity of the~outflow. 


\section{Method}
We followed the growth of dust grains and dust-driven wind by starting from the input stellar quantities, i.e.,~current stellar mass (M$_*$), Luminosity (L$_*$), effective temperature (T$_{\rm eff}$), mass-loss rate ($\dot{M}$) and~elemental abundance in the photosphere, and~by introducing seed particles, since we did not follow the nucleation process in our calculations, which is uncertain and poorly understood (e.g., \citep{Gobrecht17}). In~this work, we adopted the stellar evolutionary tracks computed with the FUNS code \citep{stra06,cris09}, freely available on the web pages of the FRUITY database\endnote{\url{http://fruity.oa-abruzzo.inaf.it}.} \citep{cris11,pier13,cris15,cris09}.
The calculation of dust growth with dust-driven wind requires solving a system of ordinary differential equations (e.g., \citep{FG06, ventura12, nanni13, Nanni14}). The~description used here built upon the one adopted in~\cite{nanni13, Nanni14} and~was improved to describe the inner part of the CSE, before~the onset of a dust-driven wind, where SiC is formed following \citep{che92}. In~particular, the~model successfully reproduced the grain size distribution of SiC presolar grains found in meteorites \citep{cris20}.
Below, we outline the theoretical description adopted in this~work. 
\subsection{Stellar Evolutionary~Tracks}\label{tracks}
 We considered the evolution of stars with the initial mass\endnote{We refer to ``initial mass'' as the mass of the star on the main sequence.} and metallicity typical of C-stars populating the Milky Way (Z $\simeq$ Z$_\odot$ $\simeq$ 0.014), the~LMC (Z $\simeq$ 0.008), SMC (Z $\simeq$ 0.003) and~Galactic Halo (Z $\simeq$ 0.001). Note that for stars with initial masses between 2 and 3 M$_\odot$, the~mass at the beginning of the TP-AGB phase almost coincides with the initial mass. The selected metallicity for SMC C-stars was slightly lower than the one usually adopted in the literature (Z $\simeq$ 0.004), but~represented the most suitable value among the grid of models available in FUNS. On~the other hand, we expected very similar results for the two metallicities. For~this study, we selected 1.5, 2 and 3~M$_\odot$, for~$Z\ge 0.003$, which provide the bulk of C enrichment in our Galaxy and are representative of the population of C-stars in the MCs (e.g., \citep{Dellagli15, Dellagli15b, Pastorelli19, Pastorelli20}), while we considered 1.3 and 1.5~M$_\odot$ at $Z\simeq 0.001$, for~the C-stars in the Galactic Halo, which are expected to be older than those in the MCs and our Galaxy. We additionally considered 2 and 3~M$_\odot$ at Z = 0.001, which can be representative of dusty C-stars in low-Z star-forming dwarf galaxies \citep{Boyer15, Boyer17}. In~models with $M>1.5$~M$_\odot$, the TDU is more efficient than at lower masses. On the other hand, the~temperatures is not high enough to activate H burning at the base of the convective envelope (the so-called hot bottom burning; see e.g.,~\citep{ventura11,kakka}), as~in more massive stars (M/M$_\odot \ge4$). 
 
We refer to~\cite{cris09,cris11} for a detailed description of the FRUITY models.
As a rule of thumb, we found that the dredge-up becomes more efficient for decreasing metallicities. 
This is a direct consequence of the stronger TPs that occur at low Z and of the following expansion of the layer above to work off the energy surplus released by thermonuclear runaways. This causes a quick drop of the H-burning shell temperature, which is no longer able to provide the required entropy barrier to prevent the envelope penetration. Moreover, at~low metallicities, the~paucity of CNO isotopes in the envelope implies a larger temperature to efficiently trigger H burning in the shell (thus, a deeper mass coordinate and, as~a consequence, a~larger TDU efficiency). M = 3 M$_\odot$ with Z = 0.003, however, represents an exception to the aforedescribed rules. This is due to the fact that at metallicities \mbox{Z $\sim$ (0.002 $\div$ 0.005)}, models with mass M $\sim$ 3~M$_\odot$ experience a transition phase, showing physical properties typical of intermediate-mass stars (see Figure~3 
 of \citep{bis10}).
Those stars have a definitely larger H-exhausted core, this fact implying a larger
compression of the H-exhausted layers. As~a consequence, ignition conditions for the 3$\alpha$ process are attained earlier, making the period between two TPs shorter. Consequently, the~TDU efficiency of this model decreases with respect to similar stellar masses with higher metallicity. Another consequence of a larger core mass is the increased surface luminosities L$_*$. The~surface temperature increases as well, but~this is mainly related to the larger stellar compactness (necessary to compensate the paucity of CNO isotopes, which act as catalysts during the H burning). As~a matter of fact, at~low Z, AGB stars show larger surface temperatures with respect to the corresponding metal-rich~equivalents.

The mass-loss rate of FRUITY models depends on both the surface temperature and luminosity. The~mass loss is calibrated on the observed $\dot M$-Period relation in Galactic giants (see \citep{stra06} for details). The~Period is derived from the M$_K$-Period relation proposed by \citep{whi03}. 
To date, no metallicity dependence has explicitly been included, even if different initial compositions or element ratios (e.g., C/O) were indirectly taken into account from the variations in the surface temperature and luminosity (via the calculation of appropriate opacities; see \citep{Cri07}). 

\subsection{Seed~Nuclei}
Grain formation in the CSE of TP-AGB stars is a two-step process. First, small solid particles (seed nuclei) are formed. Later on, grains accrete through the addition of molecules from the gas phase.
Modelling the complex phase of nucleation is beyond the scope of the paper, and the number of seed nuclei was chosen a priori in our model. Such a parameter mainly determines the grain size, once the input stellar parameters are given.
Since the seed particle abundance depends on the details of the nucleation process and on their composition, different seed particle abundances can be assumed for different grain species.
In this work, the~seed particle abundance of SiC grain was selected in order to reproduce the size distribution of SiC presolar grains, as~discussed in \mbox{\citet{cris20}}. On the other hand, the~seed particle abundance of amC grains has been constrained in a series of works in order to reproduce the infrared and optical photometry of C-stars in the \mbox{MCs \citep{nanni16,Nanni18, nanni19, Nanni19b}}. In~such a framework, the~number of seed nuclei for amC was selected to be proportional to the carbon excess, under~the assumption that seed particles are mainly composed of carbon, which is not directly linked to the initial metallicity of the star. 
 In this paper, we explored the possibility that amC grains accrete on seed nuclei whose abundance is proportional to the metallicity, as, for~example, TiC \citep{vanLoon08}. This implies that the number of seed nuclei scales with the metallicity rather than with the carbon excess.
 The choice of the seed particle abundances is summarised in Table~\ref{tab:parameters}.

\end{paracol}
\nointerlineskip
\begin{specialtable}[H]\setlength{\tabcolsep}{5.1mm}
\widetable
\caption{Reactions and input parameters selected for the calculation of dust~condensation.}
\begin{tabular}{lllll}
\toprule
\textbf{Dust Species} & \textbf{Reaction}	& \boldmath{$\alpha_i$}	& \boldmath{$\epsilon_{\rm s, i}$} & \textbf{Optical Constants}\\
\midrule
AmC		& C$_2$H$_2$ $\rightarrow$ 2C(s)+H$_2$		& $0.2$	& (
a) $\propto \epsilon_{\rm s}\times$ (C-O), $\epsilon_{\rm s} = 10^{-11}$ & \cite{Pitman08}\\
		& 		& 	& (b) $\propto \epsilon_{\rm s}\times$ Z, $\epsilon_{\rm s} = 10^{-11}$ & \\
SiC		& 2Si+C$_2$H$_2$ $\rightarrow$ 2SiC(s)+H$_2$		& $1.0$	& $10^{-15}$ & \cite{Hanner88} \\
\bottomrule
\end{tabular}
\label{tab:parameters}
\end{specialtable}
\begin{paracol}{2}
\switchcolumn

\subsection{Grain Accretion in the~CSE}
We followed the process of dust growth by computing the collision rate of the gas species on the grain surface. In~this work, we treated the formation of amC and SiC dust grains in CSEs of C-stars as~separate species. We neglected the formation of iron, which, if~present, is usually produced in very small amounts~\cite{Nanni19b}, and~of MgS, whose formation still challenges theoretical models~\cite{Zhukovska08}.
A summary of the parameters and assumptions adopted in this work is provided in Table~\ref{tab:parameters}.
For each dust species, $i$, the~variation of the grain size in time was computed by evaluating the growth and destruction rates ($J^{\rm gr}_i$ and $J^{\rm dec}_i$):
\begin{equation}\label{eq:growth}
 \frac{da_i}{dt}=V_{0,i} (J^{\rm gr}_i-J^{\rm dec}_i),
\end{equation}
where $V_{0,i}$ is the volume of the monomer of dust and $J^{\rm gr}_i$ and $J^{\rm dec}_i$ are the growth and destruction rates, respectively. In~this paper, we followed the grain accretion for amC and SiC, and therefore, two ordinary differential equations needed to be integrated. The description adopted yielded a single value of the grain size for each dust species and for each set of input stellar quantity, which changed along the TP-AGB evolution.

The growth rate, $J^{\rm gr}_i$, is determined by the slowest step of the dust growth reaction provided by the minimum among the collision rates of the gas species, $j$, called ``rate-determining species'', on~the grain surface of the dust species, $i$:
\begin{equation}
 J^{\rm gr}_i= s_i\alpha_i\min\Big\{\frac{1}{s_j} n_j v_{th, j}\Big\}, 
\end{equation}
where $s_j$ is the stoichiometric coefficient of the species $j$ in the grain formation reaction and $s_i$ is the one of the dust species. Note that in order to form $s_i$ monomers of dust, $s_j$ molecules/atoms will be needed. The coefficient $\alpha_i$ is the sticking coefficient, which provides the probability for a molecule to stick on the grain surface; $n_j$ is the number density of molecules in the gas phase; $v_{th, j}$ is the thermal velocity of the gas particles. The~quantity $n_j$ depends on the abundance of the species $j$ in the gas phase, as well as on the density, while $v_{th, j}$ is proportional to the square root of the gas temperature (see~\cite{nanni13} \mbox{for details}).

The destruction rate, $J^{\rm dec}_i$, depends on the efficiency of the erosion operated by H$_2$ molecules in the gas phase that collide with the grain surface. The~destruction rate was computed from the law of mass action for gas-phase reactions, since, at~equilibrium, the~rate of the forward reaction of grain formation equals the rate of the backward reactions. The gas species providing the destruction rate is the rate-determining species (rds).
\begin{equation}
 J^{\rm dec}_i= s_i\alpha_i \frac{1}{s_j} n_{eq, rds} v_{th, rds}, 
\end{equation}
where $n_{eq, rds}$ is the number density of the rate-determining species at equilibrium \citep{nanni13} where the abundances of all the other gas species are given. 

Despite the formation path of amC grains not yet being fully understood, we here refer to~\cite{che92}, according to whom amC grains can accrete below a gas temperature of 1100~K where the formation of PAHs becomes efficient. Therefore, to~compute amC accretion, a~fully efficient dust growth without any destruction process is considered below such a \mbox{threshold \citep{FG06}}. We also checked the stability of amC grain formation against sublimation. The~formation of SiC grains was calculated according to Equation~(\ref{eq:growth}). In~case chemisputtering is fully efficient, such a destruction process is more effective than dust sublimation operated by the heating from stellar photons, since SiC dust is a highly refractory compound. We note, however, that the efficiency of chemisputtering may be reduced significantly under certain conditions (e.g., as for silicate dust around oxygen-rich \mbox{stars) \citep{nanni13}}.
The initial size of grains was selected to be $a_0=1$~nm. 
In the specific case of the CSE of C-stars and in the framework of the description adopted here, SiC dust grains are formed before amC~dust.
\subsection{Dust-Driven~Wind}
If enough momentum is transferred to the dust grains from stellar photons, the~outflow is accelerated via a dust-driven wind. In~C-stars, the dust species driving the outflow acceleration is amC dust. On the one hand, the~opacity of SiC and MgS around the peak of the stellar radiation ($\sim$1~$\upmu$m) is negligible with respect to the one of amC dust. On the other hand, the~amount of SiC formed is not sufficient to drive a wind through scattering even though large grains formed (as for solar metallicity), while MgS is always formed too far from the stellar surface to drive the acceleration \citep{Zhukovska08}. 
Assuming spherical symmetric outflow, no drift velocity between gas and dust and~neglecting the contribution of the gas pressure, we have that the stationary acceleration for the dust-driven wind is \mbox{given by:}
\begin{equation}\label{eq:vel}
 v \frac{dv}{dt}=-\frac{GM_*}{r^2}\Big(1-\Gamma\Big).
\end{equation}

The quantity $\Gamma$ is the ratio between the radiation pressure and the gravitational pull of the star:
\begin{equation}
 \Gamma=\frac{L_*}{4\pi c G M_*}\kappa,
\end{equation}
where $\kappa$ is the opacity of the medium in cm$^2$/g \cite{nanni13}. The~outflow is accelerated via a dust-driven wind if $\Gamma>1$.
The initial expansion velocity, $v_i=4$~km~s$^{-1}$, corresponds to the lowest value observed for C-stars in the Galaxy \citep{Schoier13, Ramstedt14, Danilovich15} and is consistent with the assumption for the piston velocity in hydrodynamic simulations (e.g., \citep{Bladh19}). The~velocity was kept constant in the case the outflow was not accelerated, and our description yielded the lowest observed expansion velocity. In such a case, we assumed that some other mechanism, such as pulsation, was responsible for the outflow acceleration. 
\subsection{Structure of the~Envelope}
In order to calculate the growth rate of each dust species, we needed to compute the gas density and temperature~profile.
\subsubsection{Temperature~Profile}\label{temperature_prof}
The temperature profile was computed by following \citep{lucy76} from the condensation zone of the first dust species formed in the CSE (SiC dust) as adopted in several works in the literature \citep{FG06, ventura12, nanni13}.
The gas temperature profile is given by:
\begin{equation}\label{eq:Tout}
T_{\rm outer}(r)^4=T_{\rm eff}^4\left[W(r)+\frac{3}{4}\tau_{\rm L}\right],
\end{equation}
where $W(r)$ is a term that takes into account the dilution of the radiation with the distance from the star,
$W(r)=\frac{1}{2}\left[1-\sqrt{1-\left(\frac{R_*}{r}\right)^2}\right]$,
where $R_*$ is the stellar radius, and $\tau_{\rm L}$ is obtained by integrating the equation:
\begin{equation}\label{tauL}
\frac{d\tau_{\rm L}}{dr}=-\rho\kappa \left(\frac{R_*}{r}\right)^2. 
\end{equation}

For the inner part of the CSE, from~the photosphere to the formation of SiC, the~temperature profile is described as:
\begin{equation}\label{eq:Tinn}
 T_{\rm inner}(r)=T_{\rm eff} \Big(\frac{r}{R_*}\Big)^{-\alpha_{\rm T}},
 \end{equation}
where $R_*$ is the stellar radius and $\alpha_{\rm T}$ a parameter calculated by linking the outer and the inner temperature profiles at the SiC condensation zone, $R_{\rm cond}$, viz. $T_{\rm inner}(R_{\rm cond})=T_{\rm outer}(R_{\rm cond})$.
\subsubsection{Density~Profile}
The density profile is described by the time-averaged description for the shock-extended zone \citep{che92}:
\begin{equation}\label{Eq:rho_inn}
 \rho_{\rm inner}(r)=\rho_0\times \exp{\int_{R*}^{r} -\frac{(1-\gamma_{\rm shock}^2)}{H_0(r^{\prime})}dr^{\prime}}, 
\end{equation}
where $\rho_0$ is the density at the photosphere and $\gamma_{\rm shock}$ is the shock strength. The~quantity $H_0(r)$ is the static scale height, as~described in~\cite{che92}:
\begin{equation}\label{Eq:H0}
 H_0(r)=\frac{kT(r)r^2}{\mu m_{\rm H} G M_*},
\end{equation}
where $k$ is the Boltzmann constant, $T(r)$ the temperature profile described in Section 
 \ref{temperature_prof}, $\mu$ the mean molecular weight, $m_{\rm H}$ the mass of hydrogen atom and $G$ the gravity constant.
The quantity $H_0(r)$ was numerically integrated to derive the density at each radius once the temperature profile, $T(r)$, is provided by Equations~(\ref{eq:Tout}) and (\ref{eq:Tinn}).

If the outflow is accelerated via a dust-driven wind at a distance $R_{\rm acc}$ following the formation of amC dust, the~density profile is described by a stationary wind profile from the acceleration zone outwards:
\begin{equation}
\rho_{\rm outer}(r)=\frac{\dot{M}}{4\pi r^2 v(r)},
\end{equation}
where $v(r)$ is obtained by integrating Equation~(\ref{eq:vel}).

The parameter $\gamma_{\rm shock}$ was initially set to a value of $0.9$ and was kept constant in the case the outflow was not accelerated via a dust-driven wind by employing this initial value. If~the outflow was accelerated, $\gamma_{\rm shock}$ was recomputed in order to match the inner and the outer density profiles by satisfying the condition $\rho_{\rm inner}(R_{\rm acc})=\rho_{\rm outer}(R_{\rm acc})$. The~procedure was reiterated until convergence was achieved. In~some cases, the~value of $\gamma_{\rm shock}$ was modified during the iterative procedure even if the final value attained did not produce a dust-driven wind (i.e., if the outflow was initially accelerated with a value $\gamma_{\rm shock}=0.9$, but~did not accelerate with the iterated value of $\gamma_{\rm shock}$).

Given that the formation of SiC occurs before the one of amC dust, where this latter species is responsible for the outflow acceleration, modelling the inner part of the envelope by taking into account the effect of shocks is particularly relevant to model the formation of SiC grains, including presolar grains. In this regard, the~aforementioned prescriptions were adopted in \citet{cris20} in order to reproduce the size distribution of presolar SiC grains.

\section{Results and~Discussion}\label{comp}
\subsection{Comparison between Model Predictions and~Observations}
We considered diagnostic plots in order to investigate the effect of metallicity on dust formation.
We based our analysis on the comparison between our theoretical trends with the latest analysis of the IRS Spitzer spectra of C-stars of our Galaxy and of the \mbox{MCs \citep{Sloan16}}.
From an observational point of view, two main diagnostics were employed to analyse midinfrared spectra and investigate the condensation process around C-stars as a function of the stellar parameters: (a) the equivalent width of the absorption feature of acetylene at 7.5~$\upmu$m; (b) the SiC feature strength at 11.3~$\upmu$m with respect to the continuum.
Alternative to the 7.5 $\upmu$m absorption feature, the~3.1 and 3.8 features were also analysed in~\cite{vanLoon08}.
We discuss the predicted trends between the SiC dust fraction and of the acetylene injection rate as a function of the metallicity and of optical depth, which is considered as a proxy for the [6.4]--[9.3] 
 colour. For~such a comparison, we refer to Figures~5 and 9 
 of~\cite{Sloan16} for SiC and acetylene, respectively. 
The value of the optical depth at 1~$\upmu$m ($\tau_1$) is obtained from:
\begin{equation}\label{tau1}
 \tau_1=\int_{R_{\rm cond}}^{\infty} \kappa(1\, \mu m, r) \rho_{\rm d}(r) dr,
\end{equation}
where $R_{\rm cond}$ is the radius at which the first dust species condenses (SiC for C-stars), $\kappa(1\, \upmu\text{m}, r)$ is the opacity profile at 1~$\upmu$m and $\rho_{\rm d}(r)$ is the dust density profile. As~a consequence, $\tau_1$ depends on the amount of dust formed, the~density profile and the condensation radius. Similar values of $\tau_1$ can be obtained for different combinations of stellar~parameters.

We here show how the acetylene content in the gas phase (i.e., the acetylene injection rate) is expected to change as a function of $\tau_1$ for different metallicity values and initial stellar masses. The~acetylene content was estimated from the available free carbon in the CSE (not locked in CO molecules or in dust grains):
\begin{equation}
 \frac{dM(\rm C_2H_2)}{dt}= \frac{\dot{M}}{2} \frac{X_H}{m_H} {m_{C_2H_2}} (\epsilon_c-\epsilon_o-f_c \epsilon_c), 
\end{equation}
where $\rm X_H$ is the hydrogen mass fraction, $\rm m_H$ the mass of hydrogen, $\epsilon_c$ and $\epsilon_o$ the total carbon and oxygen abundances with respect to hydrogen and~$\rm f_c$ the number of carbon nuclei locked into dust grains with respect to the total number of carbon nuclei initially available in the atmosphere.

Furthermore, we analysed the expansion velocities obtained in our models by comparing our predictions with the values derived from the observations in the Milky Way and in the MCs \citep{Loup97, Schoier13, Ramstedt14, Danilovich15, Groenewegen_etal16}, as well as with hydrodynamic simulations \citep{Mattsson10, Eriksson14, Bladh19}.

\textls[-15]{In the following, we present the results of our analysis for two choices of seed particle abundances: (a) proportional to the carbon excess or (b) proportional to the metallicity.}
\subsection{Grain~Size}
For both choices of seed particle abundance in this work, either built up starting from acetylene or proportional to metallicity, such as TiC clusters, the~simulations performed at Z = 0.008, 0.003 yielded amC grain sizes $<0.06$~$\upmu$m, which reproduced the photometry of C-stars in the MCs from the optical to the midinfrared bands in combination with the selected optical properties \citep{nanni16, Nanni18, nanni19, Nanni19b}.
The seed particle abundance of SiC was instead kept fixed to a value of $\epsilon_{\rm s, SiC}=10^{-15}$, able to reproduce the grain size distribution of presolar SiC grains \citep{cris20}, peaking around 0.2--0.3~$\upmu$m when combined with a sticking coefficient of $\alpha_{\rm SiC}=1$. The~maximum SiC grain size attained by our simulations decreased for decreasing metallicity, down to $<$0.1 $\upmu$m for Z = 0.001, since the available amount of Si to form SiC in the gas phase decreases.

\subsection{Acetylene in the Gas Phase and Dust~Condensation}

The acetylene content reflects how much of the available carbon remains in the gas phase rather than being consumed during the formation of dust grains (see the reaction path in Table~\ref{tab:parameters}). 
The analysis of midinfrared Spitzer spectra of C-stars highlighted a dependence of the acetylene in the gas phase with metallicity (\citep{Sloan16}, Figure 9). Specifically, the~equivalent width of the acetylene absorption feature at 7.5~$\upmu$m became wider for decreasing metallicity values for the same [6.4]--[9.3] colour.
In the left panel of Figure~\ref{c2h2}, we show the acetylene injection rate as a function of $\tau_1$ for the different metallicity selected for two representative initial stellar masses, 2 and 3~M$_\odot$, and assuming the seed particle abundance to be proportional to the carbon excess. For~each track with a given stellar mass and metallicity, we evaluated the average $\tau_1$ and C$_2$H$_2$ injection rate by generating \mbox{1000 random} ages for each star selected along the TP-AGB phase and~interpolating the C$_2$H$_2$ rate between the model results along each~track.
\begin{figure}[H]
\includegraphics[scale=0.6]{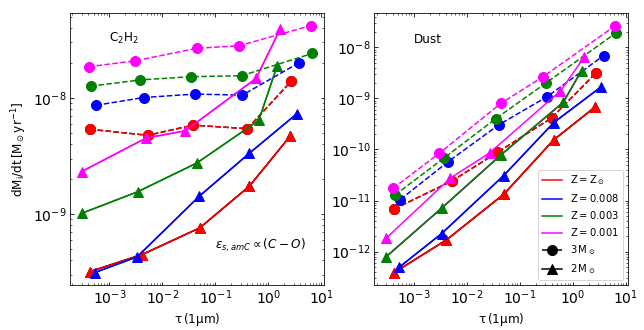}
 \caption{Injection rate of C$_2$H$_2$ molecules ({left} panel) and of dust ({right} panel) as a function on $\tau_1$ for different metallicity values and initial stellar masses, as indicated in the figure legend. The~seed particle abundance for amC dust is assumed to be proportional to the carbon excess (see Table~\ref{tab:parameters}).}
 \label{c2h2}
 \end{figure}

For a selected metallicity value, the~acetylene content increased with $\tau_1$ due to the increase of the mass-loss rate and of the carbon excess. For~a given value of $\tau_1$, the~acetylene content increased for lower metallicity values consistently with the observed trend for both the initial stellar masses: 2~M$_\odot$ and 3~M$_\odot$. This trend was due to the increase of the the carbon excess in the CSE for decreasing metallicity values at a given stellar mass produced by a higher TDU efficiency or~by an increase of the mass-loss rate. As~discussed in Section~\ref{tracks}, for 3~M$_\odot$ at $Z=0.003$, the TDU was less efficient than for the higher metallicity values, but~this was compensated by a larger mass-loss rate due to the higher surface luminosity. 
The acetylene injection rate largely differed in models characterised by different masses (but with the same metallicity). In~the most extreme case for Z = 0.008 around $\tau_1\sim 10^{-3}$, such a difference was nearly up to two orders of magnitudes. The~acetylene content also showed a steeper rise for stellar tracks of 2~M$_\odot$.
The large differences in the acetylene injection rate obtained in our models for 2~M$_\odot$ and 3~M$_\odot$ at all the metallicity values may explain the large scatter observed in the 7.5~$\upmu$m~width.

A larger amount of carbon atoms in the CSE is also associated with higher dust production, which remains however low enough (with condensation fractions typically $\lesssim$0.5) to allow acetylene to increase for decreasing metallicity. As~shown in the right panel of Figure~\ref{c2h2}, for~a given stellar mass, higher dust production rates are needed at low metallicities to obtain the same value of $\tau_1$ (see Equation~(\ref{tau1})). For~the same reason, at~a given metallicity, higher production rates are needed for larger stellar masses. This effect was produced by the higher effective temperatures of low-Z stellar tracks and larger luminosity for higher initial stellar mass moving the condensation zone outwards in the CSEs (see also Equations~(6) and (7) of \citep{Bressan98}).

\subsection{SiC Mass~Fraction}
The analysis of midinfrared spectra also highlighted a dependence of the dust feature strengths with respect to the continuum as a function of the stellar parameters (\citep{Sloan16}, Figure 5).

In Figure~\ref{sic}, we show the SiC dust fraction as a function of $\tau_1$ for different metallicity values and for two representative initial stellar masses: 2 and 3~M$_\odot$. The simulated data points were obtained by a similar procedure as in the plots previously shown for the C$_2$H$_2$ injection rate. For~a given value of $\tau_1>10^{-2}$, we clearly predicted an increase of the SiC mass fraction with metallicity due to the increase of the Si content at the stellar surface, which allows the SiC to be formed in larger amounts. The~SiC mass fraction also tended to reduce for increasing values of $\tau_1$, which was produced by an increase of amC dust due to subsequent TDU episodes.
A similar trend was also found in other theoretical works in the literature, in~which different stellar evolutionary tracks were used \citep{nanni13, Ventura16}. Our findings were qualitatively in good agreement with the observed~trend.
\begin{figure}[H]
\includegraphics[scale=0.5]{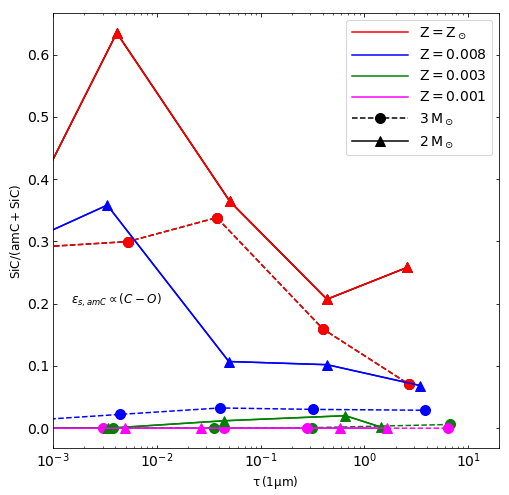}
 \caption{The mass ratio of SiC dust as a function of initial stellar mass and metallicity as indicated in the legend. The~seed particle abundance for amC dust is assumed to be proportional to the carbon excess (see Table~\ref{tab:parameters}).}
 \label{sic}
 \end{figure}
Observations from Spitzer spectra showed the SiC feature strength to increase more steeply with the [6.4]--[9.3] colour at solar metallicity, and it decreased thereafter, while the relation was flatter at lower metallicity values. Even though the results in Figure~\ref{sic} may suggest a similar qualitative trend, we needed to confirm the predicted trend by computing the spectra and the feature strength as a function of the [6.4]--[9.3] colour.
\subsection{Outflow Expansion~Velocities}

We predicted the wind speed of the outflow through a dust-driven wind. 
The comparison between observations and model predictions is shown in Figure~\ref{vexp}, where the expansion velocity is plotted against the luminosity. The~distribution of points in the plots was obtained by employing the same method as for the plots in the previous sections. Besides~the values derived for our own Galaxy in the top-left panel \citep{Schoier13, Ramstedt14, Danilovich15}, we show in the top-right panel the expansion velocities of four C-stars in the LMC observed with the Atacama Large Millimeter Array (ALMA) \citep{Groenewegen_etal16} ($Z\sim1/2$ of solar) together with three C-stars in the thick disc \citep{Lagadec12} ($Z\sim 1/3$ of solar). Thick disc stars are also shown in the bottom-left panel, since their typical metallicity is higher than the SMC one ($Z\sim 1/4$ of solar), but lower than the LMC one. In~the bottom-right panel, we also plot the expansion velocity of three C-stars in the Galactic Halo or Sagittarius Stream with an expected metallicity of 1/10 solar \citep{Lagadec12}.
\begin{figure}[H]
\includegraphics[scale=0.75]{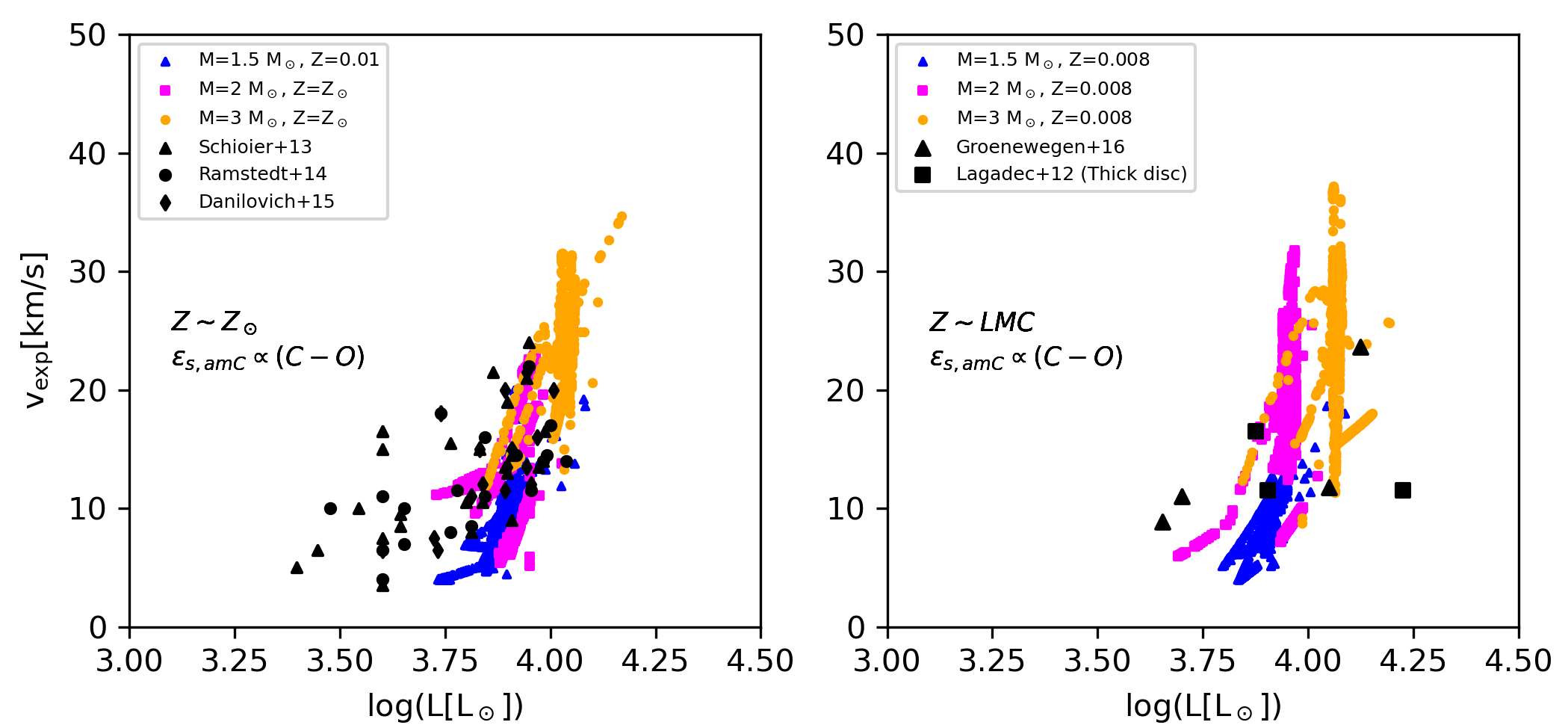}\\
\includegraphics[scale=0.75]{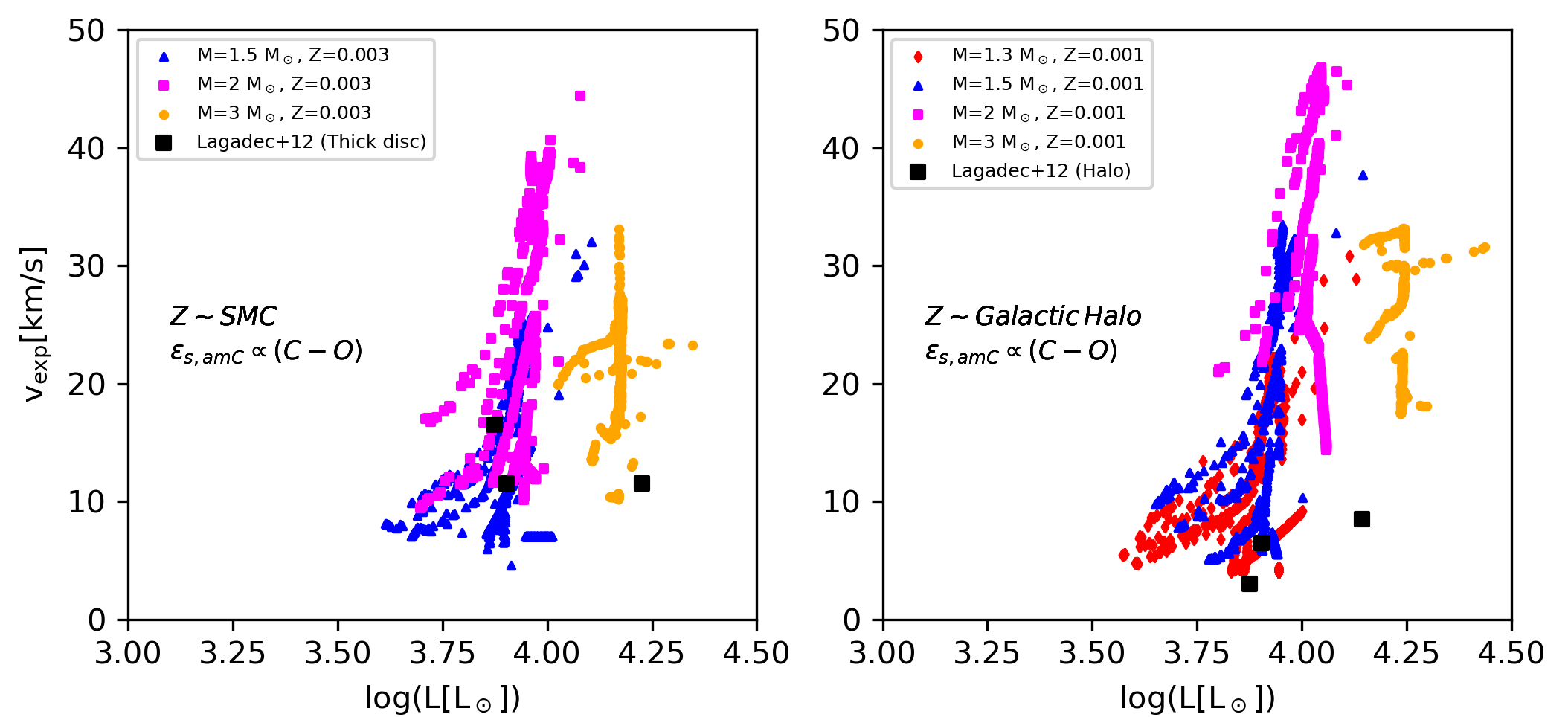}
 \caption{Comparison between observed and predicted expansion velocity vs. luminosity for different metallicity values mentioned in the panel for the selected tracks in the legend (coloured symbols). Black points represent the quantities derived from the observations of C-stars, as~mentioned in the legend. The~seed particle abundance for amC dust is assumed to be proportional to the carbon excess (see Table~\ref{tab:parameters}).}
 \label{vexp}
 \end{figure}
The observed expansion velocities were compared with the predicted models at $Z\simeq Z_\odot$, $Z=0.008$, $Z=0.003$ and $Z=0.001$, respectively, and masses of 1.5, 2 and 3~M$_\odot$. In~the top-left panel showing the solar metallicity case, we plot the case with $M=1.5$~M$_\odot$ with $Z=0.01$, since, at this stellar mass, models with $Z \geq Z_\odot$ did not undergo the C-rich phase. The~expansion velocities of the stars in the Galactic Halo were compared with the calculations for $M=1.3$~M$_\odot$, since those stars are likely older than those in the~MCs.

A comparison between expansion velocity and mass-loss rates at metallicity values representative of our Galaxy (Z = Z$_\odot$, 0.01, 0.02) and masses of 1.5, 2 and 3~M$_\odot$ is shown in the Appendix 
 \ref{appa}.
The comparison between the predicted and observed expansion velocities against the mass-loss rates of C-stars in our Galaxy \citep{Loup97, Schoier13, Ramstedt14, Danilovich15} (Figure~\ref{vexp_sun}) allowed us to select the value of the sticking coefficient for amC dust that best reproduced the observed expansion velocities ($\alpha_{\rm amC}=0.2$). We adopted this method given the uncertainties in the measurements of the sticking coefficients for which different values have been employed in the literature (e.g., \citep{FG06, Mattsson08, nanni13, Sandin20}). For~SiC, we selected a sticking coefficient of $\alpha_{\rm SiC}=1$ on the basis of the experiments of SiC sublimation growth \citep{Raback99} and~adopted in \citet{Ferrarotti02}. We adopted the same value in all the calculations of dust growth at different metallicity values throughout this work.
As shown in Figure~\ref{vexp}, the~expansion velocities against luminosity were generally well reproduced by models at varying metallicity. At~the lowest metallicity value, the~observed expansion velocities were below $10$~km~s$^{-1}$ and possibly regulated by a combination of pulsation and dust-driven wind \citep{McDonal18}. 
 For stellar tracks with 3~M$_\odot$, the expansion velocity tended to be lower for $Z<0.008$, because~of the higher effective temperature in low-Z models, which delayed the condensation of amC in the CSEs (see Equation~(\ref{eq:vel})) and was not counterbalanced by an increase of the carbon excess for decreasing metallicity.
For stellar tracks with 1.5 and 2~M$_\odot$, the opposite trend was found: the expansion velocity increased for decreasing metallicity values. For~these stellar masses, the parameter determining the increase in the expansion velocity was the carbon excess, which increased for lower metallicity values.
The results presented were in qualitative agreement with the ones obtained with hydrodynamic calculations that found that the expansion velocity increased for larger values of the carbon excess and luminosity and~for lower values of the effective temperature (e.g., \citep{Bladh19, Eriksson14, Mattsson08}). Different from the hydrodynamic calculations, however, we applied our description of dust growth and wind dynamics to stellar evolutionary tracks, where all the stellar parameters changed simultaneously.

As illustrated in Figure~\ref{vexp}, the~available observations suggested that expansion velocities decrease with the metallicity value. If~confirmed, such an observed trend would be consistent with the theoretical predictions obtained for the 3~M$_\odot$ models rather than with 1.5 and 2~M$_\odot$. However, due to the lack of statistical samples at metallicity lower than solar, it was not possible to conclusively confirm the observed trends. Furthermore, lower expansion velocity in the Galactic Halo can be also ascribable to a stellar mass smaller than the typical one of the MCs.
To address the current limitations set by the observations, it is desirable to obtain CO-based wind speed measurements for samples at different metallicities, from~supersolar down to 0.1 solar or below if possible, that span a larger and better sampled range in luminosity, as well as mass. Equally important, accurate measurements of luminosity and mass-loss rate require fitting the SED of C-stars and accounting for nonisotropic effects due to asphericity; direct measurement of both the initial metallicity (from elements unchanged by dredge-up) and current envelope composition are required for all individual stars as no samples are monometallic; likewise, direct measurement of both the initial mass (from cluster membership or spatio-chemo-kinematical association with stellar populations) and current mass (from pulsational analysis or binarity) are required for individual stars. All of these are attainable to some extent for well-chosen samples of objects and with dedicated effort.

\subsection{Seed Particle~Abundance}
The results obtained were tested for two choices of the seed particle abundance: either scaled with the carbon excess or to the initial metallicity (Table~\ref{tab:parameters}). We found that this choice did not significantly affect the results. 
This implied that it is not possible to clearly define the nature of seed particles formed in the CSE of C-stars either as composed mainly by carbon or a metallicity-related compound on the basis of the diagnostics considered here.
The invariance of the results with the choice of the seed particles was mainly due to the choice of the parameter $\epsilon_{\rm s}$ (see Table~\ref{tab:parameters}), which affected the condensation fraction of amC dust and the onset of the dust-driven wind. Specifically, $\epsilon_{\rm s}=10^{-11}$ yielded small amC grains ($<$0.06~$\upmu$m) able to reproduce the SED of C-stars in the MCs for the choice of the seed particle dependence. Indeed, the~variation of the seed particle abundance $\epsilon_{s, amC}$ produced by setting it proportional either to the carbon excess or the metallicity did not produce a substantial variation of the condensation fraction of amC~dust.
\section{Conclusions}
We coupled our model of dust evolution \citep{nanni13, cris20} with the stellar evolutionary tracks computed with the FUNS code~\cite{stra06,cris09}. Our theoretical calculations were able to reproduce the acetylene, dust content and~wind speed at varying metallicities reasonably well. 
The increase in the acetylene content at decreasing metallicity for a given stellar mass was mainly explained by an increase in the carbon excess due to TDU events, which was not compensated for by a proportional increase in dust production. The~wind speed of C-stars was fairly well reproduced at varying metallicity for different choices of the initial stellar mass, suitable for different systems, i.e.,~the Milky Way, MCs and Galactic Halo. However, the~limited amount of observations prevented a thorough study of the~subject.


\vspace{6pt} 



\authorcontributions{
Conceptualization, Ambra Nanni; methodology, Ambra Nanni; investigation, Ambra Nanni, Sergio Cristallo, Jacco Th. van Loon and Martin A. T. Groenewegen; writing---original draft preparation, Ambra Nanni, Sergio Cristallo, Jacco Th. van Loon and Martin A. T. Groenewegen; writing---review and editing, Ambra Nanni, Sergio Cristallo, Jacco Th. van Loon and Martin A. T. Groenewegen. All authors read and agreed to the published version of the manuscript.}

\funding{A.N. acknowledges support
from the Narodowe Centrum Nauki (UMO-2018/30/E/ST9/ 00082 and UMO-2020/38/E/ST9/00077).}
\institutionalreview{Not applicable}

\informedconsent{Not applicable}


\conflictsofinterest{The authors declare no conflict of interest.} 

\appendixtitles{yes} 
\appendixstart
\appendix

\section{Comparison between Predicted and Observed Wind Speed of Galactic~C-Stars}\label{appa}
\unskip
Figure~\ref{vexp_sun} shows the comparison between the observed \citep{Loup97, Schoier13, Ramstedt14, Danilovich15} and predicted expansion velocity vs. mass-loss rate for C-stars in our Galaxy. The~data points shown along the tracks were selected by generating 1000 random ages in the C-phase for each track and interpolating the mass loss and expansion velocity. Among~these points, a~number equal to the total observed points divided by the number of tracks considered was finally plotted. The~plots show that the expansion velocities were reasonably well reproduced by our description by selecting a sticking coefficient for amC dust of~0.2. 

\end{paracol}
\nointerlineskip
\makeatletter 
\setcounter{figure}{0} 
\@addtoreset{figure}{section}
\renewcommand{\thefigure}{A\arabic{figure}}
\makeatletter
\begin{figure}[H]
\includegraphics[scale=0.9]{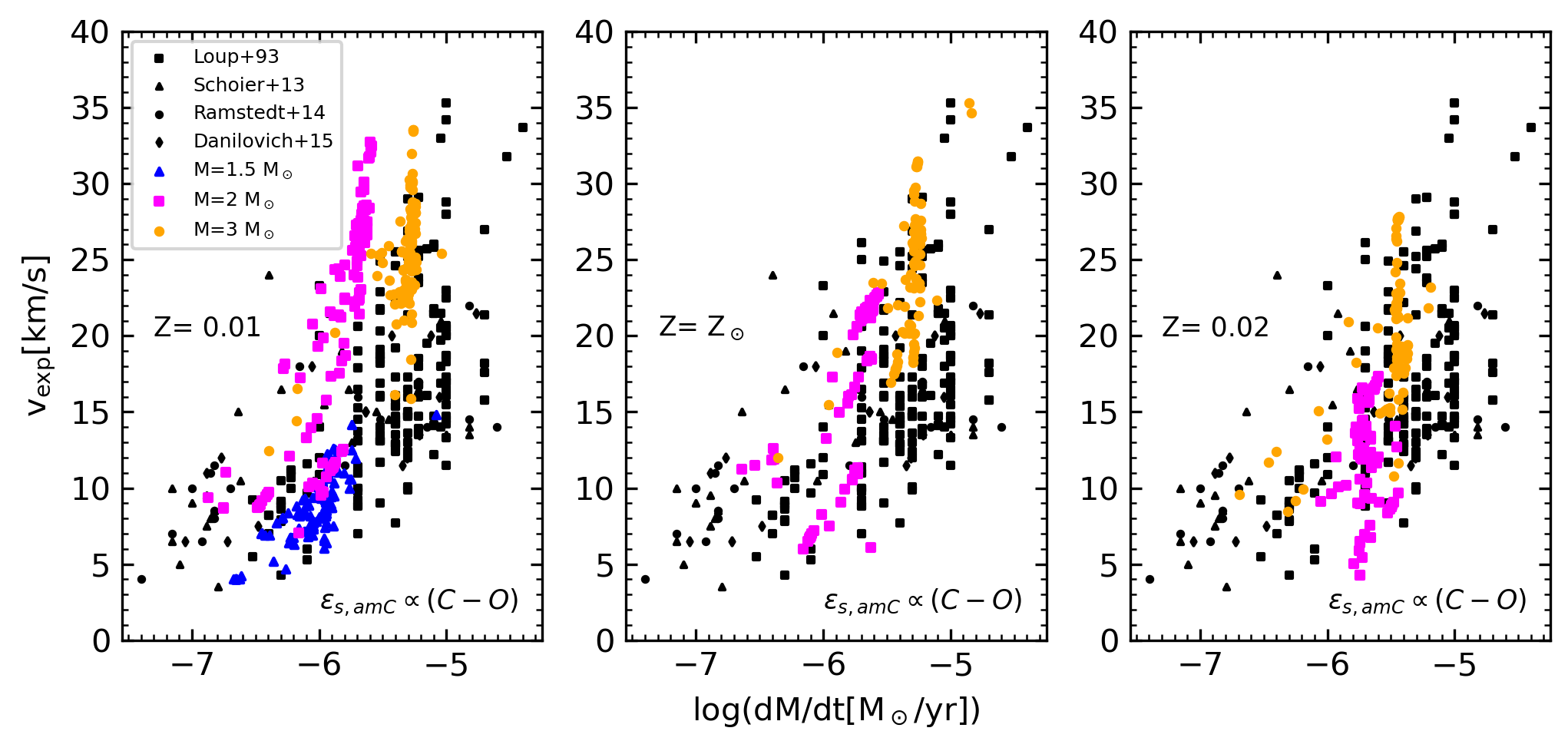}
 \caption{Comparison between observed and predicted expansion velocity vs. the mass-loss rate for metallicity representative of the Galaxy. Black points are the quantities derived from the observations. The~seed particle abundance for amC is proportional to the carbon excess (Choice (a) in Table~\ref{tab:parameters})
.}
 \label{vexp_sun}
 \end{figure}
\vspace{-3.5mm}
\begin{paracol}{2}
\switchcolumn



\end{paracol}
\printendnotes[custom]
\reftitle{References}

\end{document}